\documentclass[final,3p,times]{elsarticle}
\makeatletter
\def\ps@pprintTitle{%
  \let\@oddhead\@empty
  \let\@evenhead\@empty
  \let\@oddfoot\@empty
  \let\@evenfoot\@empty
}
\makeatother
\usepackage{float}        
\usepackage{placeins}     
\usepackage{hyperref}
\hypersetup{
  colorlinks=true,    
  linkcolor=black,     
  urlcolor=black,      
  citecolor=black,     
  pdfborder={0 0 0}   
}
\usepackage{graphicx}
\usepackage{caption} 
\usepackage{siunitx}
\usepackage{amsmath}
\usepackage{booktabs}
\usepackage{threeparttable}
\usepackage{rotating}
\usepackage{multirow}
\usepackage{arydshln}
\usepackage[normalem]{ulem}
\usepackage{color}
\usepackage[T1]{fontenc}
\usepackage[nameinlink]{cleveref}
\usepackage{float}

\newcommand{\beginsupplement}{%
        \setcounter{table}{0}
        \renewcommand{\thetable}{S\arabic{table}}%
        \setcounter{figure}{0}
        \renewcommand{\thefigure}{S\arabic{figure}}%
     }

\journal{Journal \& Under Review}

\begin{document}

\begin{frontmatter}
\title{Precipitation induced recrystallisation (PIX) in a Ti-Fe-Mo bcc-superalloy driven by lattice misfit}
\author[BHAM]{Neal M Parkes} 
\author[BHAM,CityU]{Kan Ma,\corref{cor1}}
\author[CCFE]{Ben Poole} 
\author[CCFE]{Chris Hardie} 
\author[BHAM,CCFE]{Alexander J Knowles\corref{cor1}} 
\cortext[cor1]{
Corresponding Authors. Email addresses:\\
\indent \indent a.j.knowles@bham.ac.uk (Alexander J Knowles), \\
\indent \indent kan.ma@cityu.edu.hk (Kan Ma)}
\address[BHAM]{School of Metallurgy and Materials, University of Birmingham, Birmingham, B15 2SE, UK}
\address[CCFE]{United Kingdom Atomic Energy Authority, Culham Campus, Abingdon, OX14 3DB, UK}
\address[CityU]{Department of Mechanical Engineering, City University of Hong Kong, Hong Kong, China}

\begin{abstract} 
Beta-Ti bcc-superalloys, comprising an A2 $\beta$-Ti matrix reinforced by ordered intermetallic B2 $\beta^{\prime}$-TiFe precipitates, exhibit an unusual recrystallisation that occurs with no externally applied strain (i.e. no thermomechanical processing). Thermal ageing at 750~$^{\circ}$C for 72 h results in refinement of the grain size from ~364~$\mu$m to ~30~$\mu$m. This grain refinement is driven by discontinuous precipitation of $\beta^{\prime}$-TiFe lamellae with the $\beta$-Ti matrix from grain/phase boundaries, which is associated with significant misorientation and increased dislocation density, attributed as precipitation induced recrystallisation (PIX).
\\
\\


\noindent\textbf{Graphic abstract}
\begin{center}
\includegraphics[width=12cm]{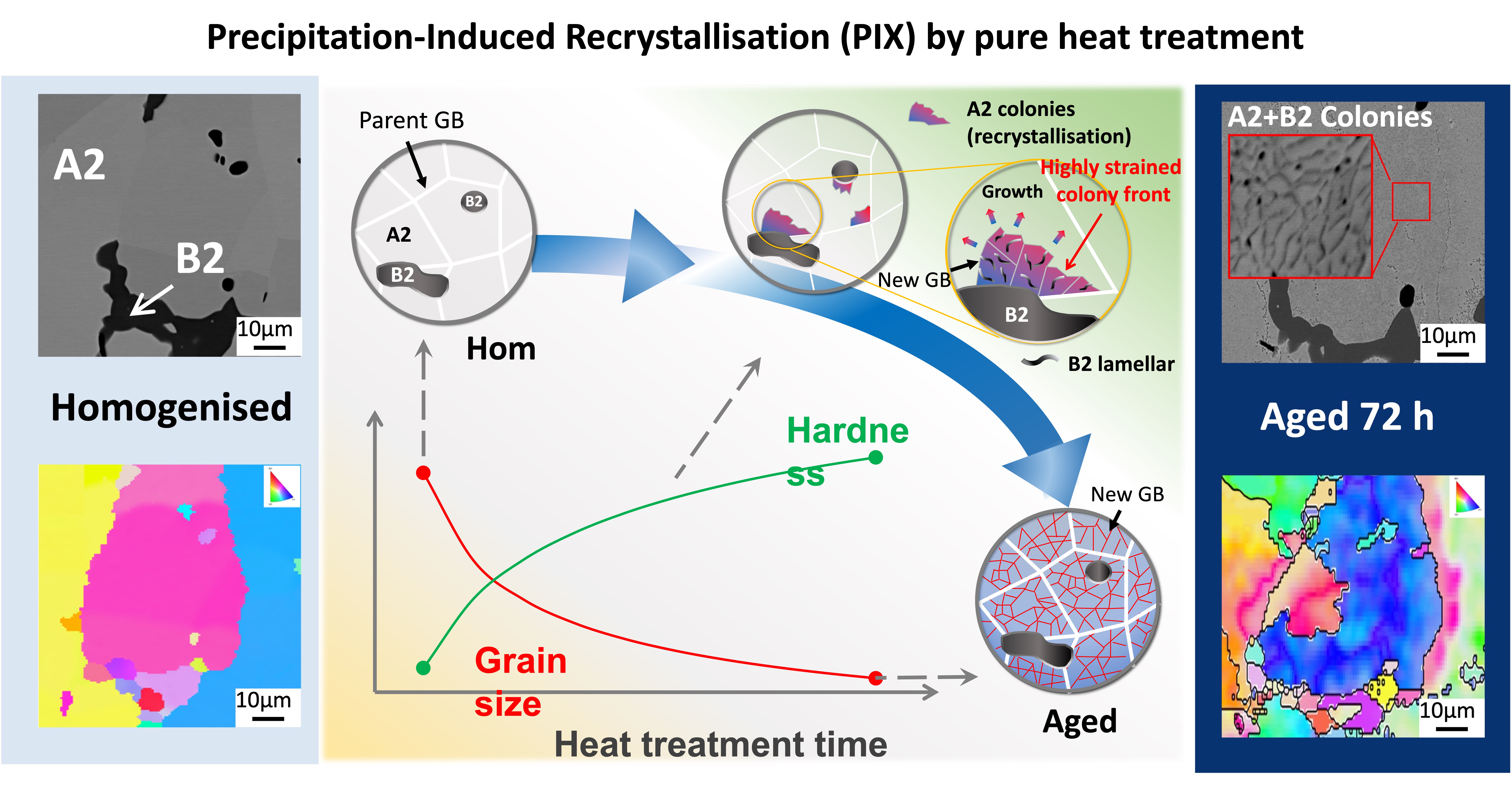}
\end{center}
\end{abstract}


\end{frontmatter}

\newpage

\section{Introduction} 
Titanium (Ti) alloys are high performance materials with exceptional specific strength \& damage tolerance used in many advance applications, particularly in aerospace, for example for jet engine compressor blades~\citep{peters2003titanium}.
The traditional approaches to improve the mechanical properties of Ti alloys 
are through a combination of (1)~grain size refinement from careful multi-step thermechanical processing $+$ recrystallisation (RX), paired with (2)~precipitation strengthening using the hexagonal close packed (hcp) $\alpha$ phase 
within a body-centred-cubic (bcc) $\beta$ matrix. \citep{zhu3,Tushar,liang,liang2}.
%

It is widely acknowledged that a refinement of $\beta$ grains play a significant role in various mechanical properties, such as strength, ductility, and fatigue resistance corrosion fatigue resistance~\citep{yu1993corrosion,bermingham2008grain}. The grain refinement necessitates the thermomechanical processing which consists of a series of hot work and heat treatments ~\citep{weiss1998thermomechanical,semiatin2020overview} with parameters precisely chosen to obtain the optimal mechanical properties and microstructure for the end-applications. Following the hot deformation in $\alpha$-$\beta$ and $\beta$ Ti alloys, subgrains form in coarse $\beta$ grains by dynamic recrystallisation and their misorientations tends to increase with strain ~\citep{semiatin2020overview, furuhara2007dynamic}.

In order to improve the operating temperature of Ti alloys $>$500~$^{\circ}$C, 
recent developments have been made to produce Ti-based "bcc-superalloys"~\citep{jones2021binary,clemens,Jovanovic,seiji,knowles2025bcc,o2023ti,zheng2018TiFe},
which, inspired by 
face-centred-cubic (fcc) Ni-superalloys with $\gamma$-$\gamma$' microstructures~\citep{reed2008superalloys},
comprise an ordered-bcc B2-TiFe intermetallic ($\beta$') in a disordered-bcc A2 titanium-molybdenum matrix ($\beta$), where molybdenum (Mo) acts as a $\beta$ stabiliser~\citep{Knowlestit,lutjering2007titanium}. Although molybdenum has limited solubility in B2-TiFe, it has full miscibility with Ti (bcc-A2) over a large region of the phase diagram~\citep{landort}. This alloy design approach induces a precipitation strengthening method similar in length scale to that seen in fcc-superalloys, and enables potent precipitate strengthening~\citep{Knowlestit}.

In the Ti-Fe-Mo system, nano-scale B2 TiFe lamellar precipitation occurs following ageing at 750~$^{\circ}$C~\citep{Knowlestit}, in contrast to cuboidal or spherical precipitates as in fcc-superalloys~\citep{reed2008superalloys} or other bcc-superalloys~\citep{ferreiros2022influence,ma2023chromium,MIRACLE2020445}. 
This is suggested to be due to the differences in 
lattice parameters of the B2 and A2 phases, of 2.98 $\pm$ 0.01 {\AA} and 3.18 $\pm$ 0.01 {\AA} respectively, giving a misfit of -6.1~\citep{Knowlestit}, which is large compared to the misfits in fcc-superalloy typically <1\%~\citep{collins2013lattice}. 
Interestingly, 
some fine A2 subgrains with about 3$^{\circ}$ misorientation were observed by TEM, separated by B2 TiFe precipitates, and it was stated "\emph{Using EBSD, partial recrystallisation was also observed as a consequence of the ageing heat treatment.}"~\citep{Knowlestit},
however, this grain refinement phenomena was not studied in detail. 
In this work, we focus on the grain microstructure under heat treatment with no externally applied strain in a Ti-Fe-Mo alloy to investigate the potential phenomena of precipitate induced recrystallisation (PIX) during ageing heat treatment, which may offer a new design strategy for grain size control in Ti alloys. 

\section{Experimental Details}
A Ti-Fe-Mo ingot was arc melted from pure (>99.9\%) element under argon gas. The ingot was repeatedly melted to ensure homogeneity.
The as-cast sample was encapsulated in a quartz ampule filled with argon and homogenised at 1170~°C for 16 hours, and water quenched. Following the same procedure, the sample was aged at 750~°C as indicated. Samples were ground and polished using a colloidal silica polishing solution. 
Vickers hardness was measured by averaging 10 indentations of 0.3~kg load after each heat treatment.
Characterisation using Scanning electron microscopy (SEM) was conducted using a JOEL 7000 with Oxford INCA energy-dispersive X-ray spectrometry (EDS). The composition of the alloy was (56.90$\pm$0.89)Ti – (21.28$\pm$4.76)Fe – (21.84$\pm$5.00)Mo (at.\%), from 5 area EDX measurements across the ingot. 
High-Resolution electron backscatter diffraction (HR-EBSD) was performed using a Zeiss EVO 10 and Thermo Fisher Apreo 2S HiVac, with Oxford Instruments Symmetry EBSD cameras and 
CrossCourt software~\citep{cross,cross2}.
%
%
%
To track recrystallization "quasi-in-situ" EBSD method was utilized following~\citep{quasi, texture}. 
Fiducial markers were used to identify Regions of interest (ROI) following homogenization. The homogenised sample was encapsulated and aged at 750~°C for 24, 48, 72, and 144 hours. Following each heat treatment, the samples were repolished for 20 minutes using OPS to remove the surface oxide layer. EBSD analysis 
was then conducted on the identified ROIs, enabling quasi-in-situ tracking of the grain evolution, similar to~\citep{quasi, texture}. 
Lattice parameters and misfit were measured by XRD using a Cu source Proto AXRD diffractometer. Diffraction spectra were collected in the 2$\theta$ range of 20-150$^{\circ}$ with a dwell time of 5 seconds, step size of 0.05 $^{\circ}$, and a total collection time of 13 hours. 
Transmission electron microscope (TEM) samples were prepared using a Focused Ion Beam (FIB) on an FEI Quanta 3D dual-beam SEM. Diffraction and Scanning TEM (STEM)/EDS was performed using a Tecnai F20 at 200 kV equipped with Oxford INCA EDS instrument. The indexation of selected area diffraction patterns of the zone axis in the matrix and precipitate was performed using the auto-index function in the software SingleCrystal based on the A2 $\beta$ Ti structure.

\section{Results \& discussion}
\subsection{Pseudo in-situ grain refinement}
\begin{figure}[htb!]
\centering
\includegraphics[width=0.4\linewidth]{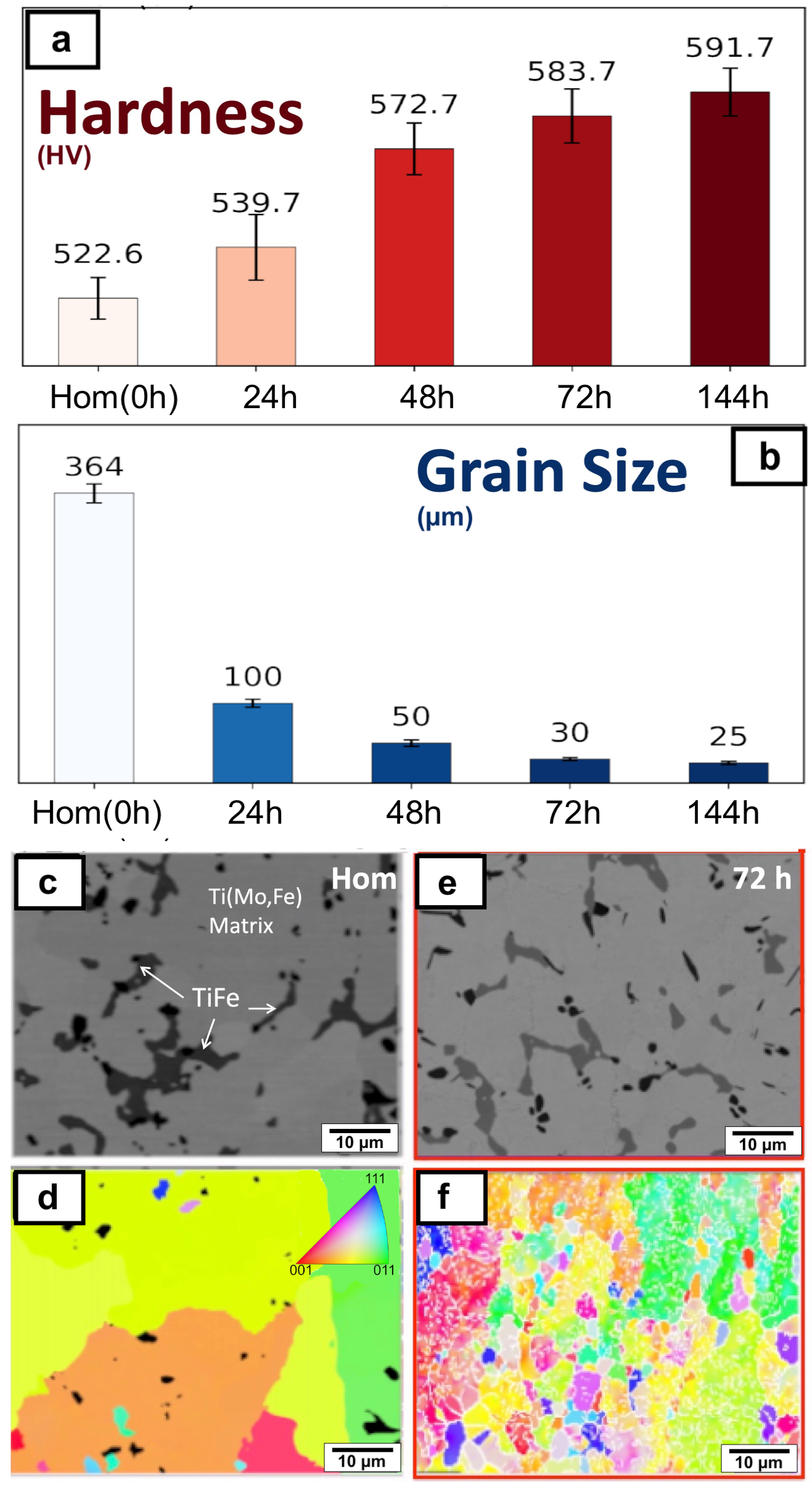}
\vspace{0pt}
\caption{Grain refinement in the Ti-Fe-Mo alloy aged without external strain: (a) Grain size and (b) hardness evolution as a function of the aging time; (c) SEM BSE micrograph and (d) EBSD IPF-x map after homogenization at 1170~$^{\circ}$C; as well as (e) SEM BSE micrograph and (f) EBSD IPF-x maps after aging at 750~$^{\circ}$C for 72h.}
\label{fig1:}
\end{figure}

The Ti-21Fe-21Mo (at.\%) alloy was homgenised and subsequently aged at 750~$^{\circ}$C for times up to 144~hours,  
Figure~\ref{fig1:}(a) and Figure~\ref{fig1:}(b), which show a progressive increase in the hardness and a decrease in average grain size by EBSD
Fig.~\ref{fig1:}(c,d) present the microstructure of the homogenised alloy along with the inverse pole figure (IPF) map, where an A2-Ti(Mo,Fe) matrix with large B2-TiFe prior dendrites observed. The microstructure after ageing for 72~h is shown in Fig.~\ref{fig1:}(e,f), with B2-TiFe prior dendrites similarly observed, however, the A2-Ti(Mo,Fe) are known to also exhibit nano-scale B2-TiFe precipitation\citep{Knowlestit}.
Curiously, given that no externally applied load was applied, e.g. thermomechanical processing, 
the grain size of the aged sample (30 $\pm$ 2~$\mu$m) was considerably smaller than that of the homogenised sample (364 $\pm$ 12~$\mu$m) (see  Supplementary Fig.~\ref{FigS1}(a,b) for grain size histograms from larger areas for enhanced statistics). 

\FloatBarrier

\begin{figure}[!htb]
\centering
\includegraphics[width=1.0\linewidth]{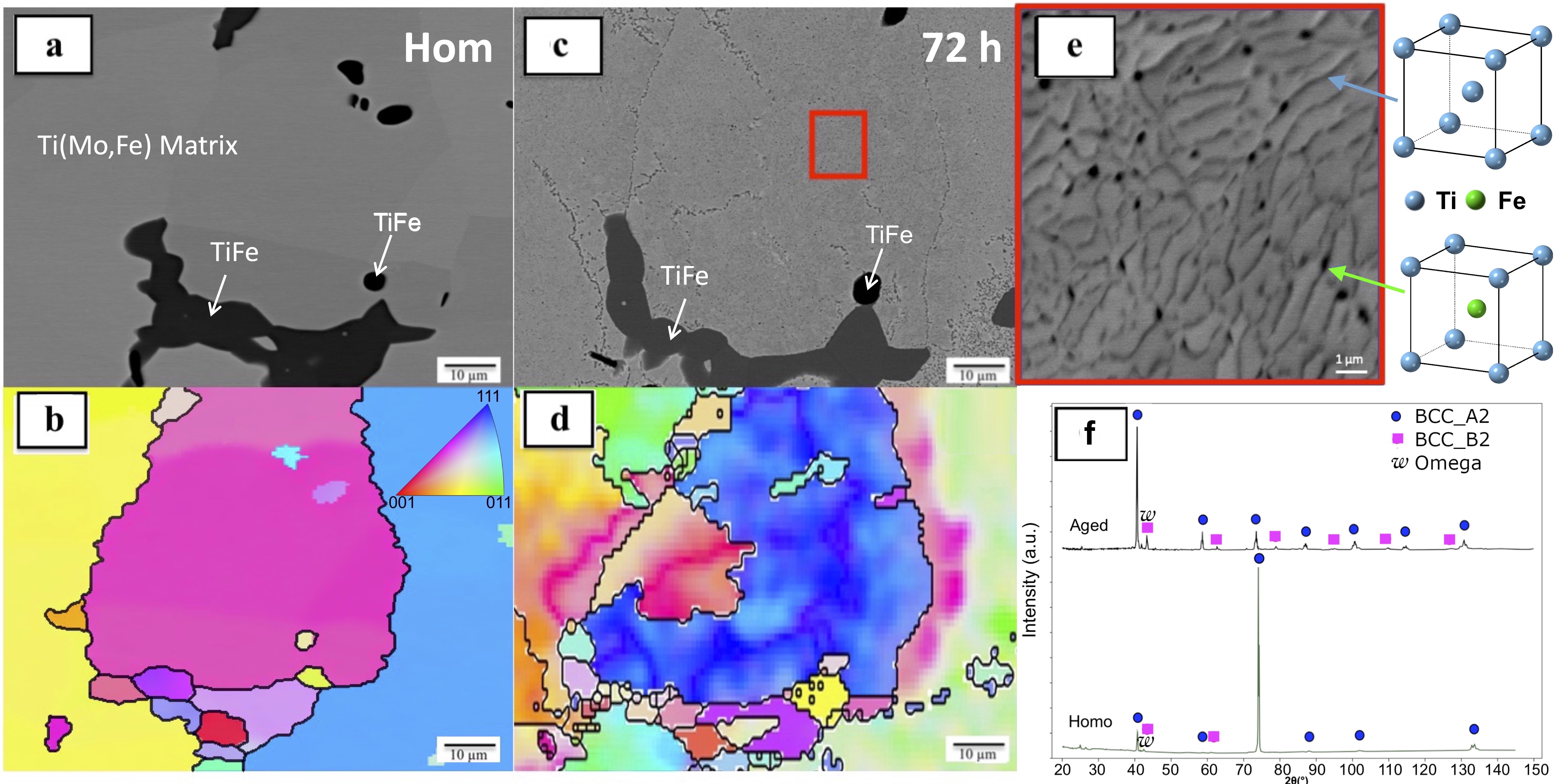}
\vspace{0pt}
\caption{Pseudo in-situ recrystallisation of TiFeMo. Microstructure evolution of a specific area by (a) BSE image and (b) EBSD IPF-x map after homogenisation at 1170~$^{\circ}$C and (c) BSE image and (d) EBSD IPF-x map after ageing for 72h at 750~$^{\circ}$C. (e) High magnification BSE of the area marked by red square in (c) showing new TiFe B2 precipitates. (f) XRD spectrum of the homogenised alloy and the aged one.}
\label{fig2:}
\end{figure}

To track the apparent grain refinement during thermal aging, a "quasi-in-situ"\citep{quasi, texture} characterization of microstructure and texture was performed tracking a specific region. From parent grains identified during homogenisation (Fig.~\ref{fig2:}(a,c)), after ageing of 72~h (Fig.~\ref{fig2:}(b,d)). 
Furthermore, several new finer grains have formed, and a strong misorientation exists within all grains, as visible by the change in IPF colour across grains in Fig.~\ref{fig2:}(d). 
Associated with the ageing, precipitation of lamellar colonies was observed (Fig.~\ref{fig2:}(e)), which were found to grow from prior grain boundaries or prior macroscopic TiFe grains (see Supplementary Fig.~\ref{fig:24}). 

\begin{figure}[htb!]
\centering
\includegraphics[width=1.0\linewidth]{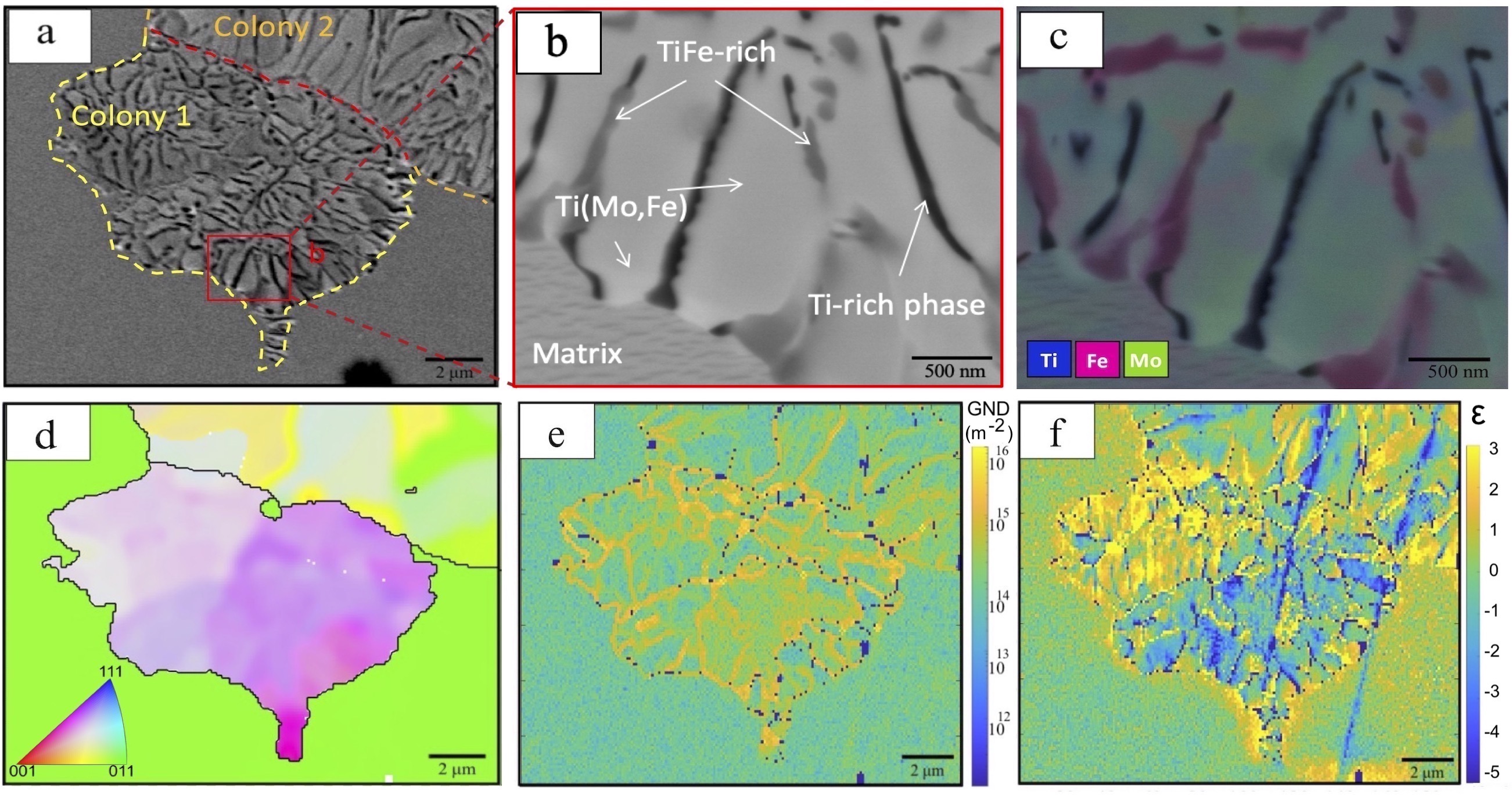}
\vspace{0pt}
\caption{Microstructure of typical lamellar colonies in TiFeMo aged at 750~$^{\circ}$C for an intermediate ageing time 24 hours. (a) BSE image showing the matrix-colonies microstructure. Matrix-colony boundaries are indicated by the yellow and orange dashed lines and the colony-colony boundary is the red dashed line. (b) A zoomed-in observation in a colony with (c) element mapping by SEM/EDS showing precipitates inside the colony. EBSD characterization in the same area as (a) showing the misorientation by (d) the IPF-x map, dislocation density by (e) GND map (GND density in dislocations m$^{-2}$) and internal strain by (f) $\epsilon_{11}$ strain map}
\label{fig3}
\end{figure}

X-ray diffraction on the alloys (Fig.~\ref{fig2:}(f)) revealed that the homogenised sample has a mostly A2 structure, except for a small peak at 42.5$^{\circ}$ attributed to the $\omega$ phase \citep{knowles,jones2021binary}. After the 72~hour ageing, where a lamellar microstructure was observed by SEM (Fig.~\ref{fig2:}(e)), the XRD patterns show two dominant sets of peaks, one related to the A2 matrix seen in the solution heat-treated sample, and a separate set from the B2 precipitates.
The presence of small quantities of A3 hcp alpha could not be clearly identified nor excluded.
The lattice parameters of the A2 structure are measured to be 
3.141~$\pm$~0.002~\AA~ in the homogenised sample, and 3.150~$\pm$~0.0011~\AA~ in the aged sample. 
The lattice parameter of the B2 precipitates in the aged sample is 2.980~$\pm$~0.086~\AA~. Using equation \ref{eq:da} \citep{Ghosh} the lattice misfit in the aged sample is -5.4 \%. 
\begin{equation}
 \label{eq:da}
   \delta = (a_{precipitate} - a_{matrix})/a_{matrix}  
\end{equation}
This misfit is consistent with the value in the Ti-Fe  binary system \citep{murray}. Such a misfit value suggests that precipitates are unlikely to be fully coherent with the matrix, as a coherent precipitate should have a misfit preferably lower than 0.1 \% \citep{Ghosh}. 
A complex semi-coherent interfacial dislocation network could therefore be formed to accommodate the lattice misfit, and induce a considerable misfit strain at the interface. As pointed out in \citep{ma2025lattice}, compared to Ni-based superalloys, the presence of large lattice misfit in bcc-superalloys will induce different microstructural evoluation and requires its own design strategies for mechanical property. In this case, the presence of strain may trigger the recrystallisation via a process known as particle-stimulated nucleation, of which the result is dependent on particle volume fraction, size, and distribution \citep{humphreys}. It is suggested that, as a result of the lattice misfit, a strain gradient builds up around the second-phase particles and induces recrystallisation (grain refinement). It is thus necessary to investigate the precipitate growth process to improve the understanding of the correlation between precipitation and recrystallisation.

The microstructure in a sample aged for an intermediate ageing time (24~h) is shown in Fig.~\ref{fig3}(a), where a lamellar colony is growing into the matrix. We remind that, before the ageing in this specific area, there was only a matrix with a similar orientation to the unconsumed matrix (area outside the colonies in Fig.~\ref{fig3}(a)). 
Fig.~\ref{fig3}(b) with the corresponding SEM/EDS map in Fig.~\ref{fig3}(c) shows three phases inside the colonies: one is the colony matrix phase rich in TiMo(Fe), while the other two are lamellar-like colony precipitates: one rich in TiFe and one rich in Ti. The SEM image analysis based on the backscattering electron contrast gray value in Fig.~\ref{fig3}(b,c) results in about 13$\pm$2\% area fraction for the TiFe-rich phase and lower than 2\% for the Ti-rich phase.
The IPF map in Fig.~\ref{fig3}(d) shows a different orientation of the growing colony from the matrix. Across both precipitate colonies, changes in orientation suggest a high level of strain within the colonies. 

Fig.~\ref{fig3}(e) presents a geometrically necessary dislocation (GND) map of the same region as Fig.~\ref{fig3}(a). The areas showing the greatest density of dislocations are located at the interface between the matrix and the growing colony, also represents the recrystallisation front. The GND values at the colony front (as indicated by the areas coloured in dotted bright yellow Fig.~\ref{fig3}(a)) exceed 1x10$^{16}$~m$^{-2}$, compared to the average GND measured to be 2.3x10$^{14}$~m$^{-2}$. In addition, a high level of GND density is also observed at the interfaces between the lamellae. The misorientation between the two colonies was measured to be a minimum of 45$^{\circ}$, representing a high-angle grain boundary.
A strain map of $\epsilon_{11}$ of the same region (Fig.~\ref{fig3}(f)) shows a high level of strain at the interface of the colony front, and precipitate-matrix interface, consistent with the dislocation density mapping (see strain maps for all directions in supplementary Fig.~\ref{FigS3}). It is worth noting that high dislocation density and strain are detected at the lamellae interfaces within the colony. 
The strain field levels around these interfaces due to precipitation can be as high as 1, which represents a strong driving force toward 
recrystallisation.  

\begin{figure}[bh!]
\centering
\includegraphics[width=0.9\linewidth]{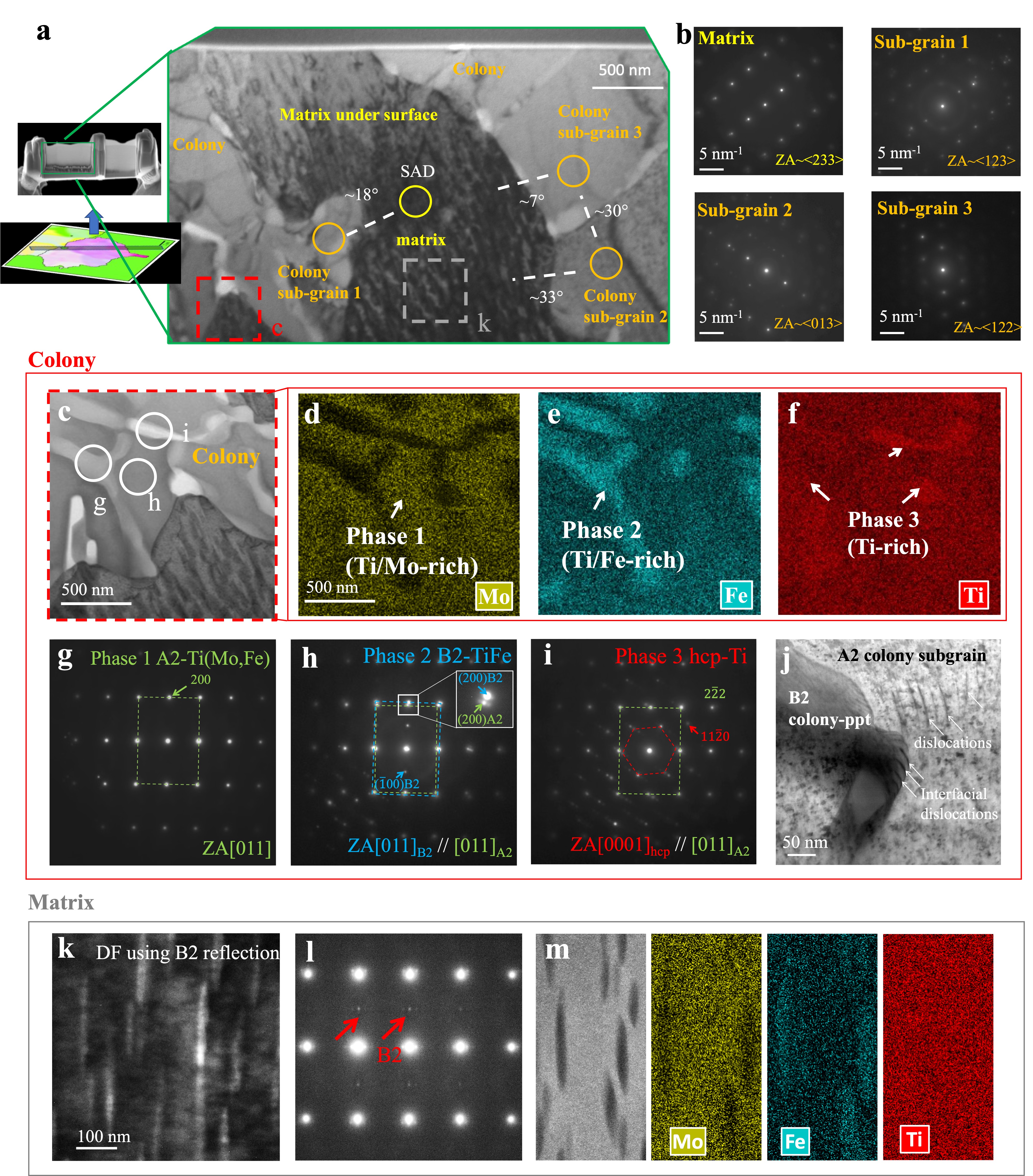}
\vspace{0pt}
\caption{(a) STEM/HAADF image of TiMoFe aged at 750~$^{\circ}$C for 24 hours showing (c-j) subgrains in colonies and (k-m) the matrix lifted out from the area in Fig.~\ref{fig3} with misorientation between them by (b) selected area diffraction (SAD) patterns. (c) STEM/EDS element mappings in (d) titanium, (e) iron and (f) molybdenum with (g,h,i) SADP of the structure of colonies and two precipitates (ppt) identified in (a). (j) STEM Bright-field images showing dislocations in a colony and interfacial dislocations between the colony and a B2 precipitate. (k) Dark-field (DF) image using a B2 superlattice reflection with (l) SAD patterns from the matrix along zone axis [011]. (m) STEM/EDS element mappings in (j) titanium (f) iron and (g) molybdenum.}
\label{fig4}
\end{figure}

\subsection{Investigation of A2+B2 microstructure in an aged alloy}
The complex precipitation-recrystallisation was further investigated using transmission electron microscopy (TEM). A TEM sample was FIB lifted out from the region in Fig.~\ref{fig3}(a). Fig.~\ref{fig4}(a) presents the overall microstructure which compromises a matrix region and two lamellar precipitate colonies, with the misorientation of the matrix and sub-grains of colonies. The misorientation has been verified at different tilt angles as exampled in Fig.~\ref{FigS4}. The determined misorientation between the matrix and colonies is indicated in Fig.~\ref{fig4}(a) and the values vary from 7$^{\circ}$ to 33$^{\circ}$. Most grain boundaries are therefore high-angle grain boundaries (HAGBs). 

Inside the colonies, two types of lamellar precipitates, a TiFe-based phase and a Ti-rich phase, are detected as shown by the STEM/EDS mapping in Fig.~\ref{fig4}(d-f), which is consistent with the SEM observation in Fig.~\ref{fig3}(b). The corresponding selected area diffraction (SAD) patterns from the different phases in Fig.~\ref{fig4}(g-i) were used to identify their structure. In Fig.~\ref{fig4}(g), the TiMo-rich colony matrix within the colony is evidenced to have a $\beta$~A2 structure, whereas the TiFe precipitates have a $\beta^{\prime}$~B2 structure, with a lattice misfit of about -6\% shown by the diffraction spot separation (inset in Fig.~\ref{fig3}(h)), and a slight rotation $\sim$1$^{\circ}$ to the A2 matrix. The lattice misfit is consistent with the XRD. The Ti-rich precipitates are indexed as $\alpha$~Ti with a hcp structure. 
The formation of $\alpha$Ti precipitate may be related to oxygen impurity but their role on the recrystallisation is considered to be minor due to their low volume fraction, $<$~2~area\% from SEM.
The diffraction patterns in Fig.~\ref{fig4}(i) show that (0001)$_{\alpha}$ $\parallel$ (011)$_{\beta}$ and [$11\Bar{2}0$]$_{\alpha}$ $\parallel$ [$1\Bar{1}1$]$_{\beta}$. This is a typical Burgers orientation relationship between $\alpha$ and $\beta$ of  \citep{burgers1934process}. There are other weak spots close to the $\alpha$~Ti spots which could be related to the doubled diffraction \citep{williams1996transmission,knowles2021tungsten}.

The STEM bright field image in Fig.~\ref{fig4}(j) shows a high density of interfacial dislocation network at the sub-grain precipitate interfaces.
Close to these precipitates, dislocations in the precipitate sub-grains are frequently observed, suggesting that these precipitates with highly strained interfaces induce a high-level deformation on the precipitate which could be the driving force for boundary movements and recrystallisation. However, within the colonies, the HAGBs are associated with B2-TiFe lamellar precipitates, which could also act to pin the growing A2 - Ti(Mo, Fe) grains. 

Outside the advancing colonies, a large number of plate-like/needle-like nano-scale precipitates were observed in the matrix as shown in Fig.~\ref{fig4}(k-m), also in the bottom left-hand corner of Fig.~\ref{fig3}(b) and bottom right-hand of Fig.~\ref{fig4}(c). The matrix precipitates have widths of approximately $\sim$~10~nm and are homogeneously distributed within the matrix. These fine precipitates are $\beta^{\prime}$ B2-TiFe and have a cube-to-cube orientation relationship with the matrix <100>$_{\beta^{\prime}}$$\parallel$<100>$_{\beta}$, $\{100\}_{\beta^{\prime}}\parallel\{100\}_{\beta}$. By comparing their morphology and the SAD patterns, these precipitates grow preferentially on the {011} planes.
These B2 precipitates should also induce additional strengthening on the matrix. Unlike the precipitates seen in the colonies, those observed in the matrix have a particular orientation. This adds credence to the idea of a high interface strain and semi-/in-coherency of the precipitates within the grain boundary front. 

%

\subsection{Strengthening mechanisms}
\begin{figure}[ht!]
\centering
\includegraphics[width=0.8\linewidth]{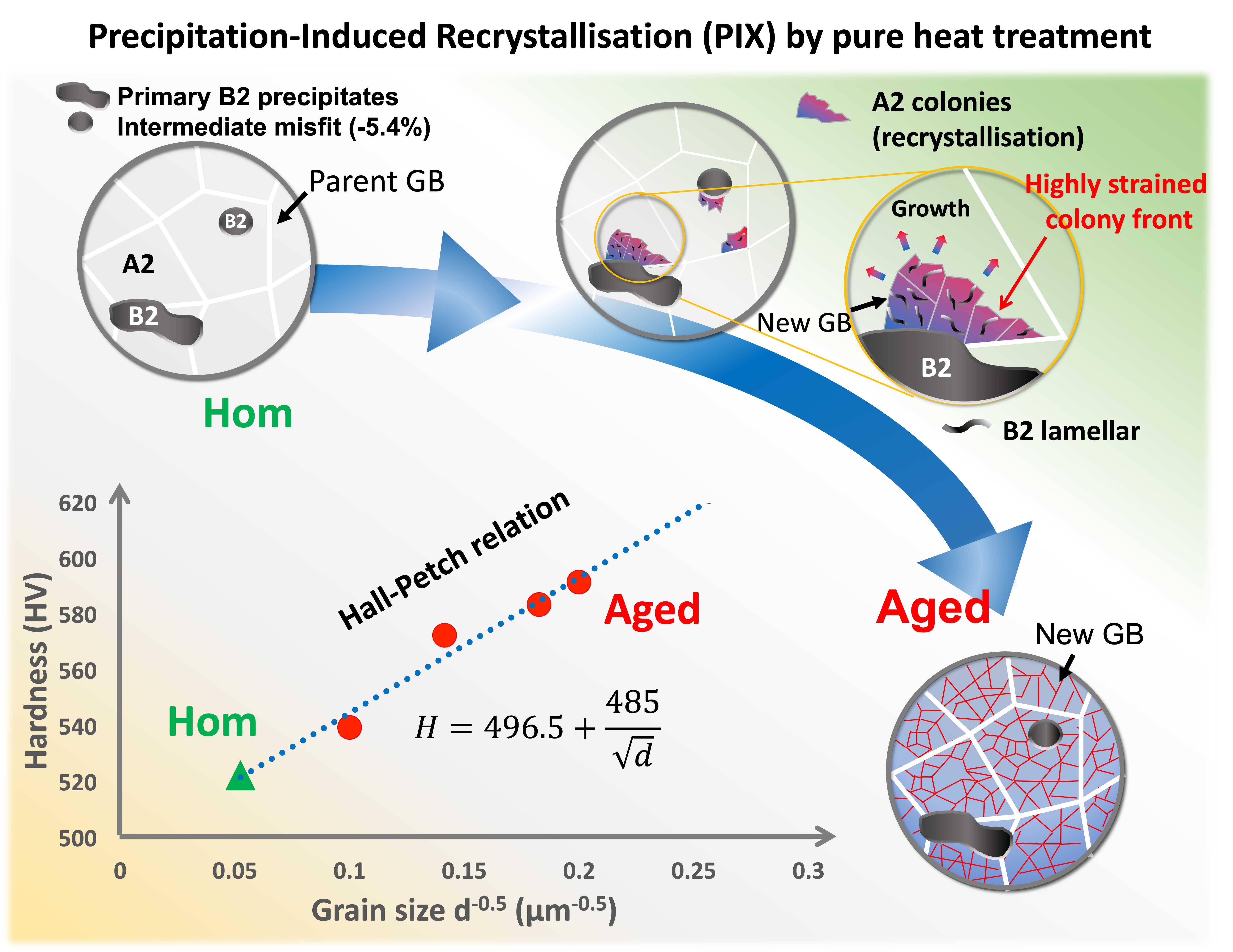}
\vspace{0pt}
\caption{Schematic illustration depicting the precipitation-induced recrystallisation (PIX) and hardening following Hall-Petch relation in a A2+B2 TiFeMo alloy during pure heat treatment.}
\label{fig5}
\end{figure}

As a result of the precipitation, colony formation and growth, new grains and sub-grains are observed, reducing the overall grain size, namely recrystallisation. As a colony forms at a prior grain boundary, it grows with a new grain orientation, while simultaneously the interface strain between the A2-Ti(Fe,Mo) and B2-TiFe phases 
drive a recrystallisation and the nucleation of sub-grains behind the colony front, which are then pinned by the B2 TiFe precipitates. We term this phenomena 'precipitation induced recrystallisation' (PIX), which offers exciting possibilities for the control of grain size and texture, without the need for externally applied deformation, \textit{e.g.} by transitional thermomechanical processing. Fig.~\ref{fig5} presents a comprehensive depiction of the microstructural evolution during ageing. Surprisingly, here we establish a compelling correlation between measured hardness and grain size, aligning closely with the well-established Hall-Petch (H-P) relation~\citep{hall1951deformation,petch1953cleavage} as expressed for hardness by:
\begin{equation}
H = H_0 + kd^{-1/2}
\end{equation}
with $k$=485~HV~$\mu$m$^{1/2}$ the overall strengthening (hardness) coefficient and $H_0$=496.5 HV the overall friction hardness. The strengthening coefficient for hardness $H$ (in MPa) can be converted into a value for yield stress $\sigma$ (in MPa) using Tabor's empirical equation $\sigma = H/3$. Thus, the strengthening coefficient for yield strength is 1585~MPa~$\mu$m$^{1/2}$, which aligns with the overall coefficient in other dual-phase Ti-based alloys (1420-3150~MPa~$\mu$m$^{1/2}$~\citep{paradkar2009validity}, 1236~\citep{sen2007microstructural},1330-2280~MPa~$\mu$m$^{1/2}$~\citep{jung1996hall}).

The dependence of yield strength on grain size through the H-P equation is a widely recognized phenomenon in single-phase polycrystalline alloys. Attempts have been made to extend the application of the H-P relationship to alloys that contain two phases and undergo undergo stress-induced martensitic transformation~\citep{paradkar2009validity,fan1993extension, sure1984mechanical}. In this work, the strengthening by PIX in the Ti-Fe-Mo alloy is shown to obey H-P relation. During the ageing, three strengthening mechanisms involves: grain boundary (GB), precipitate, and solid-solution strengthening.

Assuming that the phase boundaries show similar effects as the grain boundaries, the dislocation pile-up model to explain the H-P relationship can be extended to dual-phase alloys \citep{paradkar2009validity,cottrell1964mechanical}. The subgrains in colonies decorated by twisted TiFe laths could result in a dislocation pile-up at the grain/phase boundaries. In the completed recrystallised alloys, we consider the size of subgrains with B2 are "effective" grain size, which leads to an increase in hardness of $\Delta H_{GB}$ = 90~HV (details see Appendix). The primary B2-TiFe precipitates (average size 4~$\mu$m) are reasonably assumed to play a minor role in strengthening after ageing. The solid solution strengthening decreases with the ageing time as the Fe and Mo content in $\beta$-Ti reduces after ageing due to TiFe precipitation. The decreases of hardness is estimated to be $\Delta H_{ss}$ = -20~HV from the homogenised alloy to the 144-hour aged alloy. Finally it leads to an increase of hardness of about 70~HV in agreement with the observation, which reaffirms the pivotal role of grain refinement by B2 precipitation in augmenting the material's hardness during the heat treatment process. 

\FloatBarrier
\section{Conclusions}
To conclude, a Ti-Fe-Mo bcc-superalloy comprising a bcc A2 $\beta$-Ti matrix reinforced by ordered-bcc B2 TiFe $\beta^{\prime}$ intermetallic precipitates has been studied, which exhibits an unusual recrystallisation phenomenon, whereby ageing with no externally applied deformation results in substantial grain refinement, with the following conclusions:
\begin{itemize}
  \item 
  Ageing of the Ti-Fe-Mo alloy results in a grain size refinement from 364 $\pm$ 12~$\mu$m when homogenised, to 30 $\pm$ 2~$\mu$m after a 72~hour ageing.
  This is correlated with the formation of TiFe intermetallic precipitates, which have high misfit of -5.4\%.
  \item 
  Pseudo-in-situ EBSD observations revealed heterogeneous precipitation and growth of colonies from prior grain boundaries, with accumulation of strain and dislocations at the colony growth front, and between the A2 and B2 lamellea. 
  \item 
  TEM study within the colonies found the A2-Ti(Mo,Fe) colony matirx is separated mainly by B2-TiFe-rich precipitates, the high degree of misorientation between the sub-grains and high interfacial dislocation density suggest the B2-TiFe-rich precipitates act to pin the A2-Ti(Mo,Fe) matrix, leading to the high dislocation density and strains is observed at the matrix-colony fronts and within the colonies by EBSD.
    \item 
  The precipitation induced recrystalisation and 
  grain refinement led to an increase in the Vicker's hardness of the alloy, offering a new method for improving the mechanical properties via grain size \& texture control of alloys without need for thermomechanical processing. 
\end{itemize}

\section*{Acknowledgements}
\vspace{-3mm}
\noindent\small{This work has been carried out within the framework of the EUROfusion Consortium, funded by the European Union via the Euratom Research and Training Programme (Grant Agreement No 101052200 - EUROfusion) and from the EPSRC (grant number EP/W006839/1). Views and opinions expressed are however those of the authors only and do not necessarily reflect those of the European Union or the European Commission. Neither the European Union nor the European Commission can be held responsible for them.
A.J.K. acknowledges EPSRC University of Birmingham Impact Acceleration Account (EP/X525662/1), UKRI Future Leaders Fellowship (MR/T019174/1 \& MR/Y034155/1), Royal Academy of Engineering Research Fellowship (RF/201819/18/158). 
The authors thank Prof David Dye and Prof Ben Britton for early discussions on Ti-Fe-Mo alloys~\citep{Knowlestit}.
K.M. acknowledges support from the National Natural Science Foundation of China (52501024).
Electron microscopy was performed in the Facility for Electron Microscopy (University of Birmingham) and the Materials Research Laboratory (UKAEA), the authors thank the staff for their support \& assistance in this work.
}

\bibliographystyle{unsrt}
\bibliography{References.bib}

\hfill \\

\beginsupplement
\clearpage
\section{Supplementary Information}  
\begin{figure}[ht]
\centering
\includegraphics[width=0.75\textheight]{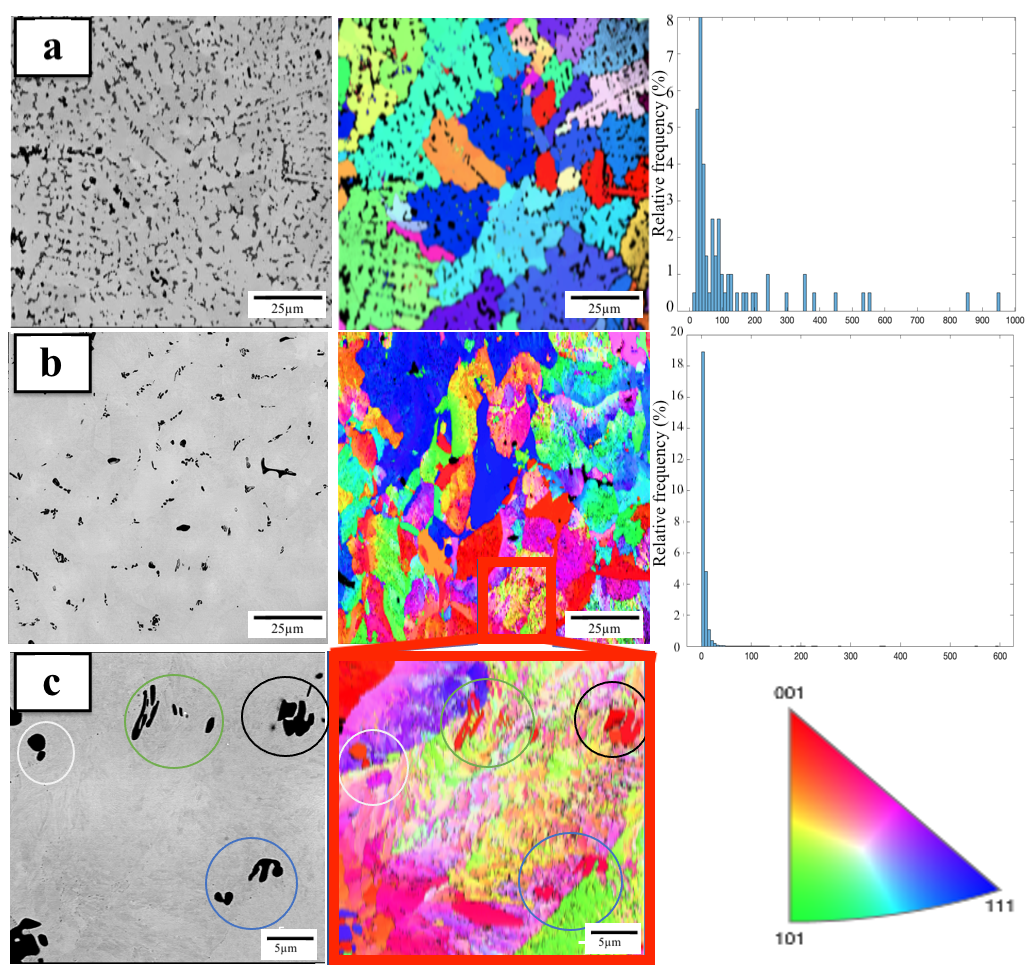}
\caption{BSE and EBSD images of the typical microstructure of TiFeMo (a) Homogenised at 1170~$^{\circ}$C along with grain size histogram. (b) Aged 750~$^{\circ}$C 72 hours BSE and EBSD of the same area along with grain size histograms. (c) Zoomed in EBSD of area marked by the red square in figure (b) to show a high degree of strain in individual grain.}
\label{FigS1}
\end{figure}

\begin{figure}
\centering
\includegraphics[width=\linewidth]{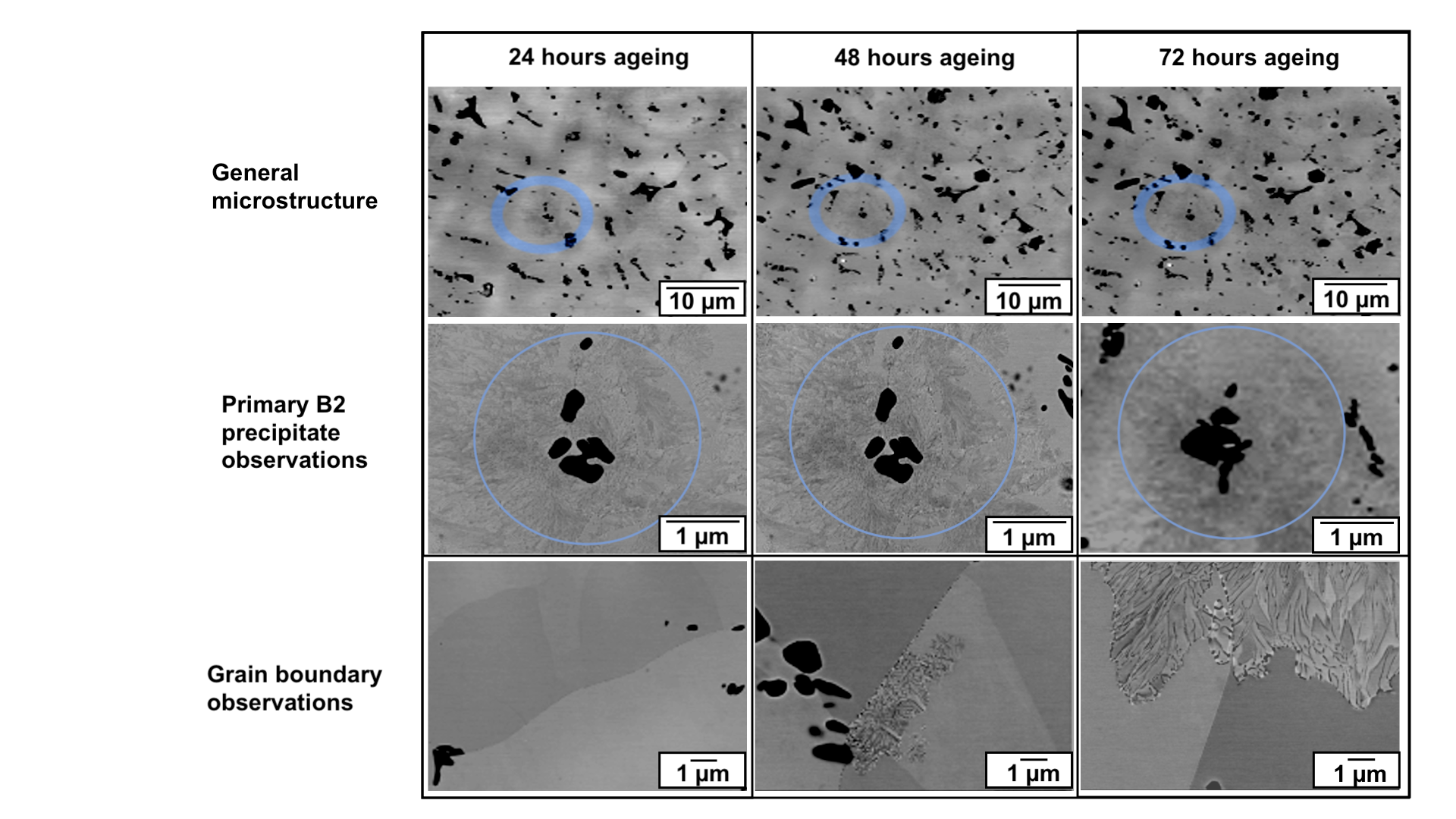}
\vspace{0pt}
\caption{Pseudo in-situ observations of the colony growth. General microstructure of a specific area in TiFeMo aged for (Column 1) 24 hours, (Column 2) 48 hours, and (Column 3) 72 hours showing the colony growth. The blue circle shows a typical primary B2-TiFe precipitate development over a 72-hour heat treatment. To further elucidate the growth of colonies BSE images of grain boundary morphology are shown over the same time period.}
\label{fig:24}
\end{figure}


\begin{figure}
\centering
\includegraphics[width=0.6\linewidth,height=0.4\textheight]{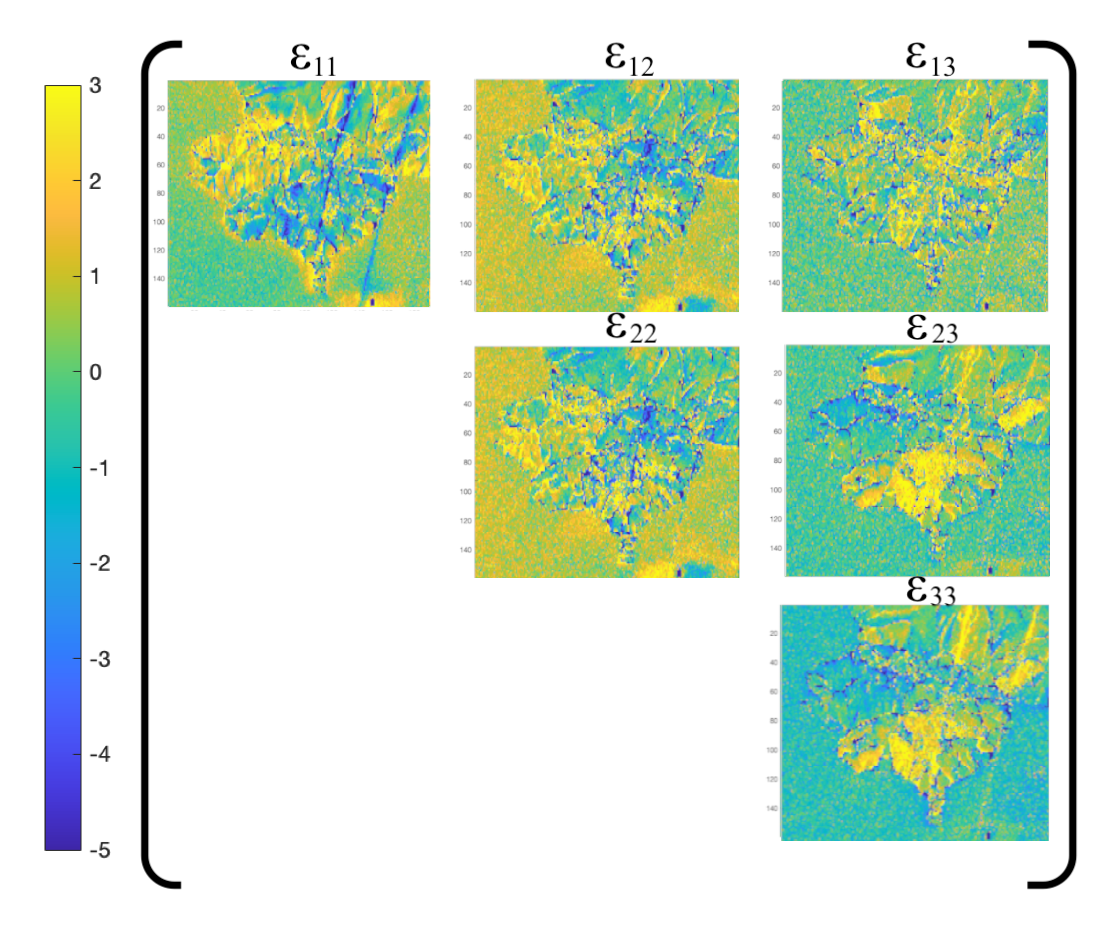}
\vspace{0pt}
\caption{HR-EBSD analysis showing strain maps of the colony as shown in Fig.~\ref{fig3}}
\label{FigS3}
\end{figure}

\begin{figure}
\centering
\includegraphics[width=1\linewidth]{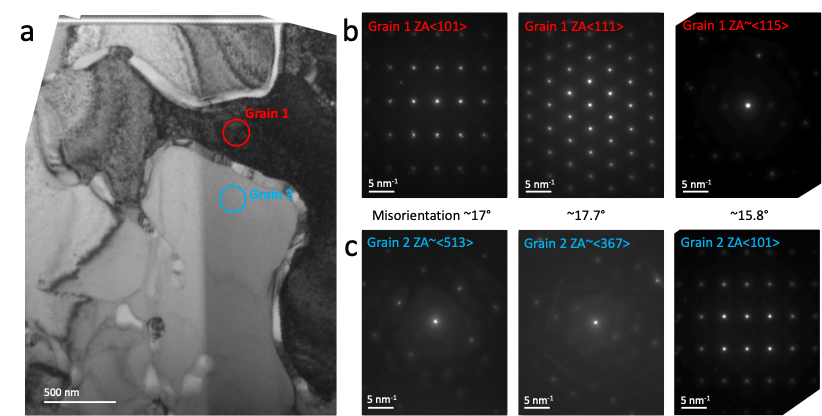}
\vspace{0pt}
\caption{(a) Bright-field TEM image of a TEM lamella from TiFeMo (750~$^{\circ}$C, 24~h) showing the misorientation of two sub-grains in a colony. Diffraction patterns of both sub-grains are given in (b) and (c) for three tilt angles resulting in similar misorientation angles.}
\label{FigS4}
\end{figure}

The solid-solution strengthening was calculated using the model in ~\cite{zhao2019modelling}:
\begin{equation} \tag{S1}
\Delta \sigma_{ss} = (\Sigma_i {B_i}^{3/2} X_i)^{2/3}
\end{equation}
where $\sigma_{ss}$ is the increase in yield strength, $B_i$ are solid-solution strengthening coefficients for solute $i$ and $X_i$ are solute atomic fraction. Here $B_{Fe}$=1715~$MPa~{at.}^{-2/3}$ and $B_{Mo}$=575~$MPa~{at.}^{-2/3}$ from ~\cite{zhao2019modelling}. From SEM/EDS, $X_Fe$=0.14 and $X_Mo$=0.29 in the homogenised alloy and $X_{Fe}$=0.1 and $X_{Mo}$=0.30 in the 144-hour aged alloy.
It leads to $\sigma_{ss,hom}$=580~MPa and $\sigma_{ss,aged}$=501~MPa, which is converted to hardness by $\Delta H_{ss,hom}\approx \Delta \sigma_{ss}/3$=179 HV and $\Delta H_{ss,aged}\approx \Delta \sigma_{ss}/3$=155 HV. The decrease in hardness due to reduction of s.-s. strengthening is about -20~HV.

The average size of primary B2 precipitates was about 4~$\mu$m with a volume fraction of about 5\%. The precipitate spacing $L$ was deduced approximately to be 5.2 $\mu$m. It is reasonable to assume that Orowan bypassing is the dominating precipitation strenthening mechanism for these large precipitates. The increase in shear stress was thus deduced using the formula from~\cite{jones2021binary}:
\begin{equation} \tag{S2}
\Delta \tau_{ppt} = \frac{\mu b}{L}
\end{equation}
where $\mu$=40~GPa \cite{jones2021binary} is the shear module of the matrix, $b$=0.274~nm the Burgers vector of a/2<111> dislocations. It results in $\Delta \tau_{ppt}$= $2.2$~MPa. Again, the increase in hardness is converted as 
$\Delta H_{ppt}\approx \Delta \sigma_{ppt}\times3~=~(\Delta \tau \times 3\times M)\times3$
=~5~HV with M=2.9 the Taylor factor for bcc materials. The B2 precipitate strengthening in the homogenised sample is thus negligeable.

Finally, the grain boundary strengthening in the homogenised alloy is negligealbe (average grain size of 360 $\mu$m). The increase in shear stress by grain boundary strengthening was deduced as from the average interspacing of B2 lath as "effective" grain size:
\begin{equation} \tag{S2}
\Delta \tau_{GB} = \frac{\mu b}{L_{B2-lath}}
\end{equation}
with $L_{B2-lath}$=360$\pm$50~nm. It leads to $\tau_{GB}$=30~MPa. Thus the increase in hardness is 90~HV. 

\end{document}